\documentclass[aps,twocolumn,nofootinbib,showpacs]{revtex4}
\usepackage{graphicx}
\usepackage{amsfonts}
\usepackage[utf8]{inputenc}
\usepackage[T1]{fontenc}
\usepackage{amsmath,amssymb,physics,graphicx,hyperref}
\usepackage{lmodern}
\usepackage{amssymb}
\usepackage{amsbsy}
\usepackage{amsmath}
\usepackage{latexsym}
\usepackage{natbib}
\usepackage{bm}
\usepackage{color}
\usepackage{float}

\begin{document}
\title{Conformal Symmetry and Non-Singular Scalar field Collapse}
\author{Mohamed Aarif A\footnote{mohamedaarif.a2023@vitstudent.ac.in}}
\affiliation{School of Advanced Sciences, Vellore Institute of Technology, \\ 
Tiruvalam Rd, Katpadi, Vellore, Tamil Nadu 632014, India}
\author{Soumya Chakrabarti\footnote{soumya.chakrabarti@vit.ac.in}}
\affiliation{School of Advanced Sciences, Vellore Institute of Technology, \\ 
Tiruvalam Rd, Katpadi, Vellore, Tamil Nadu 632014, India}


\date{\today}

\begin{abstract}
We investigate the gravitational collapse of a massive scalar field in a conformally flat, spherically symmetric spacetime within general relativity. The collapsing matter distribution is modeled using a minimally coupled homogeneous scalar field together with both perfect-fluid and dissipative matter sectors. Imposing conformal flatness through the vanishing of the Weyl tensor considerably constrains the geometry and enables the construction of exact analytical solutions. In the non-dissipative case, the field equations admit a separable conformal factor leading to a continuously collapsing configuration smoothly matched to an exterior Schwarzschild spacetime. The collapse proceeds asymptotically and does not develop a shell-focusing singularity within finite proper time. We further examine the possibility of self-similar evolution associated with homothetic symmetry. It is shown that self-similar solutions are incompatible with a perfect-fluid configuration alone, but become consistent when dissipative effects in the form of a radial heat flux are included. The resulting self-similar collapse must be matched to an exterior Vaidya spacetime and exhibits a monotonically decreasing Misner--Sharp mass due to outward energy transport. For both classes of solutions, the proper radius remains finite throughout the evolution, preventing the formation of shell-focusing singularities within the considered domain. The scalar field sector satisfies the null energy condition for the potentials studied, while the effective fluid sector exhibits violations of the null and strong energy conditions, indicating the emergence of effective exotic matter behavior. The analysis highlights the combined role of conformal symmetry, scalar field dynamics, and dissipative transport in modifying the late-time behavior of gravitational collapse and provides a possible framework for asymptotically non-singular collapse scenarios.
\end{abstract}
\maketitle

\section{Introduction}
Gravitational collapse has always been a subject of interest since the foundational work of Oppenheimer and Snyder \cite{Opp}. It describes an imploding stellar distribution which has exhausted all sources of internal energy required to maintain an equilibrium with gravity. Depending on the initial configuration, there can indeed be formation of dynamical equilibrium states, however, for a collapse without any bound, the central expectation is a formation of spacetime singularity. The Cosmic Censorship Conjecture states that such singularities are generically covered by horizons and are inaccessible to external observers \cite{penrose}. There are counterexamples to this conjecture as well, where the collapse leads to naked singularities \cite{yodz, seif, eard, christozero, new, Joshid1, Joshi2, Joshid2, Goswami, Joshi} which allow causal interaction with the exterior. Given the fact that cosmic censorship may not be universally valid, there remains a requirement for systematic investigations of physically reasonable collapse scenarios \cite{pankaj1, pankaj2}.  \\

Within General Relativity (GR), models of gravitational collapse are extensively studied \cite{Celer, Giambo, Harada, Goswami2, Dadhich, Joshid1, Joshi2, Joshid2} and the same can also be said for modified theories of gravity \cite{Goswami4, scnb1, scnb2, jar, hwang, ghosh}. Different models of collapse primarily involve different choices of matter. Scalar fields act as one such popular choice, due to their root in a Lagrangian formulation as well as their relevance in cosmology \cite{scalarcosmo}. Collapse of a scalar field, as a topic of interest, was first studied by Christodoulou \cite{christo1, christo2, christo3}, who demonstrated that depending on initial conditions, the collapsing field may either disperse or form a singularity. Follow-up investigations have confirmed this critical behavior, predominantly for mass-less scalar fields in the context of black hole formation \cite{goswami, giambo, piran, chop, brady, gund, Gundlach1}. In contrast, solutions of collapsing massive scalar fields are rare. Giambo \cite{giambo} explored the necessary conditions for singularity formation for collapsing massive scalar fields, while Goncalves \cite{gong1} investigated scalar field collapse in an asymptotically Einstein-de Sitter background. Goswami and Joshi \cite{goswami2} constructed models leading to naked singularities; Ganguly and Banerjee \cite{koyel} demonstrated that collapse may proceed even with violation of energy conditions, with implications for dark energy clustering. There are also notable investigations on scalar field collapse with a negative cosmological constant \cite{baier}, conformally flat configurations \cite{scnb3} and self-similar scalar field collapse \cite{scnbprd}. It is already accepted that scalar fields play a central role in the context of inflation \cite{Paddy} and dark energy \cite{varun, sami}. Despite extensive studies, the clustering properties of scalar field dark energy remain unclear. Massive scalar feld collapse models can provide insights into this issue. Moreover, scalar fields with suitable potentials can effectively mimic various realistic forms of matter; for example, quadratic and quartic potentials reproduce dust-like and radiation-like behavior, respectively \cite{gong2, msami}.  \\

Motivated by this, we study a model of massive scalar field collapse in conformally flat spacetime. We derive two classes of exact solutions under the assumption that the scalar field is homogeneous. For the first solution, the conformal factor is characterized by an additive function while the second one is self-similar. None of the collapsing models reach a zero-proper volume at any finite time. The non self-similar solution is non-radiating and the inclusion of a radial heat flux brings in a self-similarity. We also discuss the smooth matching of the collapsing geometries with suitable exteriors, namely, with a Schwarzschild exterior for the non-radiating solution and with a Vaidya exterior for the dissipative, self-similar solution.

\section{Conformally Flat Collapse of a Minimally Coupled Scalar Field}
We choose the spacetime metric to have a vanishing Weyl tensor, implying conformal flatness. The Weyl tensor represents the trace-free part of the Riemann curvature tensor. It encodes the free gravitational field, including tidal distortions and gravitational radiation. It is defined as
\begin{eqnarray}\nonumber
&& C_{\mu\nu\alpha\beta} = R_{\mu\nu\alpha\beta} - \frac{1}{2}\Big(g_{\mu\alpha}R_{\nu\beta} - g_{\mu\beta}R_{\nu\alpha} - g_{\nu\alpha}R_{\mu\beta} \\&&
+ g_{\nu\beta}R_{\mu\alpha}\Big) + \frac{R}{6}\left(g_{\mu\alpha}g_{\nu\beta} - g_{\mu\beta}g_{\nu\alpha}\right),
\end{eqnarray}
where $R_{\mu\nu\alpha\beta}$ is the Riemann tensor, $R_{\mu\nu}$ is the Ricci tensor and $R$ is the Ricci scalar. Our metric is conformally flat because it can be written as a conformal rescaling of the Minkowski metric, $g_{\mu\nu} = \Omega^2(x^\alpha)\,\eta_{\mu\nu}$. In four dimensions, a necessary and sufficient condition for conformal flatness is the vanishing of the Weyl tensor, $C_{\mu\nu\alpha\beta}=0$. Such spacetimes do not support independent tidal degrees of freedom or gravitational radiation and admit a high degree of symmetry that simplifies the Einstein field equations. This class is closely related to the existence of self-similar/homothetic structures in the spacetime. These are widely studied in the context of radiating and shear-free stars \cite{som, maiti, modak, bhui, patel, schafer, ivanov, herrera, ranjan}. We write the metric as                         

\begin{equation}
\label{metric}
ds^2=\frac{1}{{A(r,t)}^2}\Bigg[dt^2-dr^2-r^2d\Omega^2\Bigg].
\end{equation}
$1/A(r,t)$ acts as the conformal factor and determines the dynamical evolution of the sphere. We consider a spherical collapse of this conformally flat geometry in the presence of perfect fluid as well as a self-interacting scalar field $\phi$ minimally coupled to gravity. The action can therefore be written as
\begin{equation}\label{action}
\textit{A}=\int{\sqrt{-g}d^4x[R+\frac{1}{2}\phi^\mu\phi_\mu-V(\phi) + L_{m}]},
\end{equation}
where $V(\phi)$ denotes the self-interaction potential of the scalar field and $L_{m}$ is the Lagrangian density of the perfect fluid. Varying the action with respect to the metric yields the energy-momentum tensor associated with the scalar field, as
\begin{equation}\label{minimallyscalar}
T^\phi_{\mu\nu}=\partial_\mu\phi\partial_\nu\phi-g_{\mu\nu}\Bigg[\frac{1}{2}g^{\alpha\beta}\partial_\alpha\phi\partial_\beta\phi-V(\phi)\Bigg]. 
\end{equation}

For a perfect fluid the energy-momentum tensor is given by
\begin{equation}\label{EMT}
T_{\alpha\beta}=(\rho+p)u_{\alpha}u_{\beta}-p g_{\alpha\beta},
\end{equation}
where $\rho(r,t)$ and $p(r,t)$ are the energy density and isotropic pressure of the fluid. We restrict our considerations to a homogeneous scalar field, i.e., $\phi = \phi(t)$. The Einstein field equations can then be derived as
\begin{equation}
\label{fe1}
3\dot{A}^2-3A'^2+2AA''+\frac{4}{r}AA'=\rho+\frac{1}{2}A^2\dot{\phi}^2+V(\phi),
\end{equation}

\begin{equation}
\label{fe2}
2\ddot{A}A-3\dot{A}^2+3A'^2-\frac{4}{r}AA'=p+\frac{1}{2}A^2\dot{\phi}^2-V(\phi),
\end{equation}

\begin{equation}
\label{fe3}
2\ddot{A}A-3\dot{A}^2+3A'^2-\frac{2}{r}AA'-2AA''=p+\frac{1}{2}A^2\dot{\phi}^2-V(\phi),
\end{equation}

\begin{equation}
\label{fe4}
\frac{2\dot{A}'}{A}=\dot{\phi}\phi'=0.
\end{equation}

The homogeneous scalar field satisfies the Klein-Gordon equation
\begin{equation}
\label{wave}
\Box\phi+\frac{dV}{d\phi}=0,
\end{equation}
which, for the metric Eq. (\ref{metric}) reduces to
\begin{equation}
\label{wave2}
\ddot{\phi}-2\frac{\dot{A}}{A}\dot{\phi}+\frac{1}{A^2}\frac{dV}{d\phi}=0.
\end{equation}

Here, an overdot and a prime denote differentiation with respect to $t$ and $r$, respectively. Subtracting Eq. (\ref{fe3}) from Eq. (\ref{fe2}) we find the condition for pressure isotropy,
\begin{equation}\label{isotropy}
\frac{A''}{A}-\frac{A'}{rA}=0.
\end{equation}
This equation can be readily integrated to obtain
\begin{equation}\label{evolution}
A(r,t)=\gamma(t)r^2+\beta(t),
\end{equation}
where $\gamma(t)$ and $\beta(t)$ are arbitrary functions of time. Substituting Eq.~(\ref{evolution}) in the $G_{01}$ Eq. (\ref{fe4}), we find that $\dot{\gamma}(t)=0$, i.e., $\gamma(t)$ must be a constant. A positive value of this constant does not yield physically acceptable solutions, and hence we set $\gamma(t)=-\gamma_0$, with $\gamma_0 > 0$. The conformal factor then takes the form
\begin{equation}\label{evoluf}
A(r,t)=\beta(t)-\gamma_{0}r^2.
\end{equation}

It is worth noting that the special case $\gamma_0=0$ reduces the metric to that of a spatially flat Friedmann-Robertson-Walker spacetime. While Eq. (\ref{evoluf}) determines the radial dependence of the conformal factor, the time evolution is encoded in $\beta(t)$, which remains to be determined. In the next section we solve for $\beta$ by considering a smooth matching of the conformally flat spacetime with a suitable exterior.

\subsection{Matching of the interior space-time with an exterior Schwarzschild geometry}
For a complete description of gravitational collapse, it is essential to impose appropriate junction conditions between the collapsing interior and a suitable exterior spacetime. We begin by considering the most general spherically symmetric interior metric of the form
\begin{equation}
ds^2=N^2(t,r)dt^{2}-B^2(t,r)dr^{2}-C^2(t,r)(d\theta^{2}+\sin^2\theta d\phi^{2}),
\end{equation}
while the exterior spacetime is taken to be a Schwarzschild geometry expressed in advanced null coordinates as
\begin{equation}
ds^2=\left(1-\frac{2M}{r}\right)d\nu^2+2drd\nu-r^2(d\theta^2+\sin^2\theta d\phi^2).
\end{equation}
$M$ denotes the total mass enclosed within the boundary hypersurface $\Sigma$, and $\nu$ is the retarded time coordinate. The mass function introduced by Misner and Sharp \cite{misner} is defined as
\begin{equation}\label{misner}
m(t,r)=\frac{C}{2}(1+g^{\mu\nu}C_{,\mu}C_{,\nu})
=\frac{C}{2}\left(1+\frac{\dot{C}^2}{N^2}
-\frac{C'^2}{B^2}\right),
\end{equation}
which represents the total energy contained within a sphere of radius $C$. The continuity of the first and second fundamental forms across the boundary $\Sigma$ leads to the matching conditions \cite{santos, chan, Kolla, Maharaj}
\begin{equation}\label{mass}
M {=^{\Sigma}} m(t,r),
\end{equation}
and
\begin{eqnarray}\nonumber \label{extrinsic}
&& 2\left(\frac{\dot{C'}}{C}-\frac{\dot{C}N'}{CN}-\frac{\dot{B}C'}{BC}\right)
{=^{\Sigma}}
-\frac{B}{N}\Bigg[\frac{2\ddot{C}}{C}-\Big(\frac{2\dot{N}}{N}\\&&
-\frac{\dot{C}}{C}\Big) \frac{\dot{C}}{C}\Bigg]+\frac{N}{B}\Bigg[\Big(\frac{2N'}{N}+\frac{C'}{C}\Big)\frac{C'}{C}-\Big(\frac{B}{C}\Big)^2\Bigg].
\end{eqnarray}

These relations ensure the smooth matching of the metric and the extrinsic curvature across $\Sigma$. For the conformally flat interior spacetime, we assign the boundary hypersurface as $r=r_b , 0 <r_b <\infty$. Substituting the metric functions into the junction condition Eq. (\ref{extrinsic}), the continuity of the extrinsic curvature at $\Sigma$ yields
\begin{equation}
2\frac{\ddot{A}}{A}+2\frac{\dot{A}'}{A}-3\frac{\dot{A}^2}{A^2}-4\frac{A'}{Ar_{b}}+3\frac{A'^2}{A^2}=0.
\end{equation}

Using Eq. (\ref{evoluf}) and evaluating at $r=r_b$, the above equation can be integrated to obtain the first integral
\begin{equation}\label{1stint}
\dot{\beta}^2-4\gamma_{0}\beta = \lambda_{0} (\beta-\gamma_{0} r_{b}^2)^3,
\end{equation}
where $\lambda_{0}$ is an integration constant. For $\lambda_{0} > 0$, Eq. (\ref{1stint}) reduces to a nonlinear differential equation of the form $\dot{\beta}^2-a_{0}\beta-a_{1}{\beta}^2-a_{2}{\beta}^3-a_{4} = 0$, whose general solution involves inverse hyperbolic and elliptic functions, rendering the analysis considerably complicated. To obtain a tractable solution, we therefore restrict ourselves to the case $\lambda_{0} = 0$. In this case, Eq. (\ref{1stint}) simplifies significantly and can be integrated directly to write
\begin{equation}\label{exactevoconf}
\beta(t)=\gamma_0(t-t_0)^2,
\end{equation}
where $t_0$ is a constant of integration. Substituting Eq. (\ref{exactevoconf}) into Eq. (\ref{evoluf}), the conformal factor is explicitly written as
\begin{equation}\label{exactevoconf2}
A(r,t)=\gamma_0(t-t_0)^2-\gamma_{0}r^2.
\end{equation}

The Misner-Sharp mass for this spacetime follows from Eq. (\ref{misner}) and is given by
\begin{equation}
m(t,r)=\frac{r^2}{2A}\Bigg[\frac{r}{A^2}\Big(\dot{A}^2-A'^2\Big)+2\frac{A'}{A}\Bigg].
\end{equation}

The matching condition requires that the above expression, evaluated at $r=r_b$, is equal to the Schwarzschild mass $M$.

\subsection{Analysis of the solution and evolution of physical quantities}
\begin{figure}[h]
\begin{center}
\includegraphics[width=0.4\textwidth]{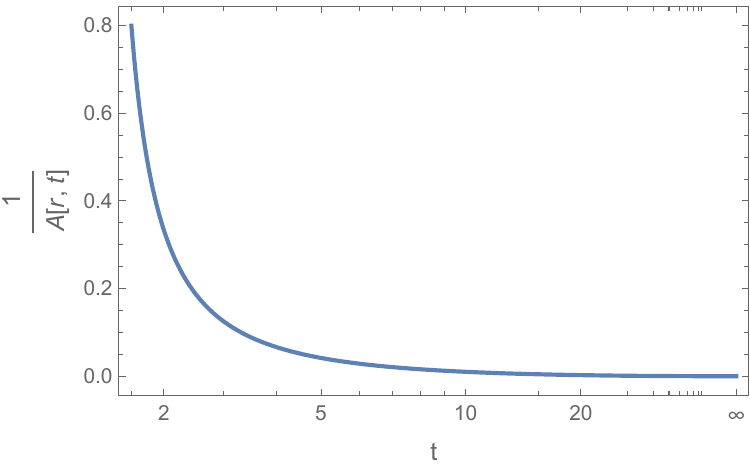}
\caption{Time-evolution of the scale factor $Y(r,t) \propto \frac{1}{A(r,t)}$ for specific values of $r$,$\gamma_{0}$ and $t_0$; the figure shows the time evolution when $t_0 < 0$ }
\end{center}
\label{fig:conformal}
\end{figure}

The scale factor $Y(r,t)$, inversely proportional to the conformal factor, can be expressed as
\begin{equation}\label{finalexactevo}
Y(r,t)=\frac{1}{\gamma_0(t-t_0)^2-\gamma_{0}r^2}.
\end{equation}                        
We plot $Y(r,t)$ as a function of $t$ in Fig. \ref{fig:conformal} for a specific value of $r < r_b$, $\gamma_{0}$ and $t_{0}$. At $t = 0$ and $r=r_b$, i.e at an initial epoch, the radius of the two-sphere has the value
\begin{equation}
Y_{initial}=\frac{1}{\gamma_0(t_{0}^2-r_{b}^2)}.
\end{equation}
Since the radius of the two sphere can never have any negative value, one must choose the parameters such that $t_{0}^2>r_{b}^2$. It is evident that the spherical collapsing body reaches a zero proper volume singularity only at infinite time, i.e. the system remains forever collapsing. This behavior holds true for any choice of the parameter $\gamma_{0}$. 

\begin{figure}[h]
\centering
\includegraphics[width=0.4\textwidth]{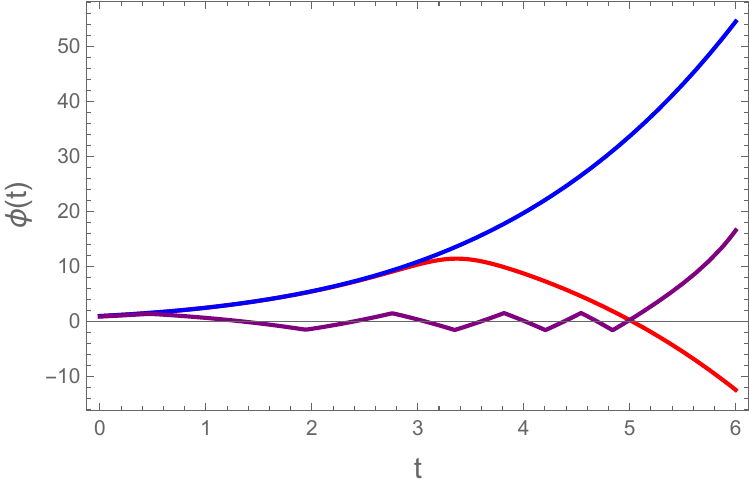}
\includegraphics[width=0.4\textwidth]{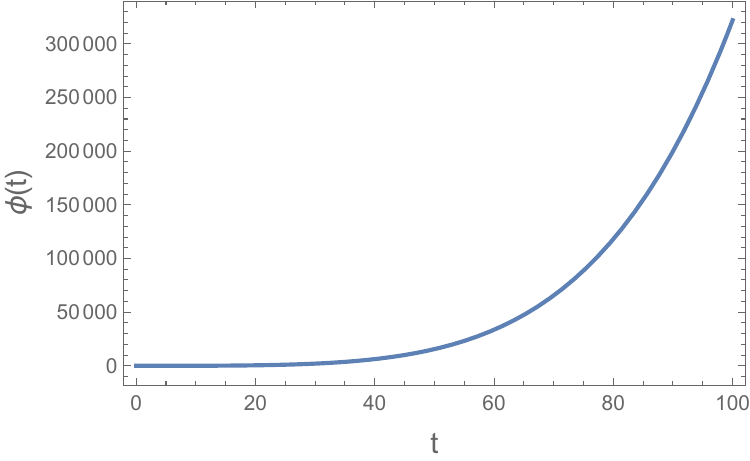}
\includegraphics[width=0.4\textwidth]{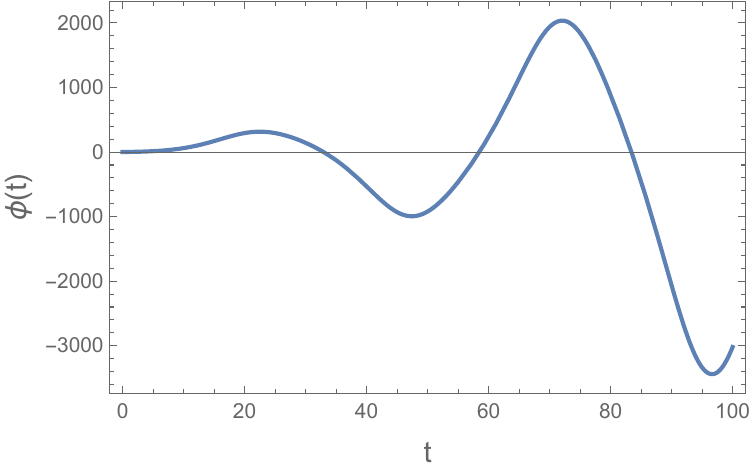}
\caption{Evolution of scalar field for three different potentials (a) $V(\phi)=a+e^{g\phi}$ (Exponential on top), (b)$V(\phi) = V_0 \log \left(1 + \frac{\phi}{\phi_0^2}\right)$ (logarithmic in middle), (c) $V(\phi) = V_0 + \frac{1}{2}m^2 \phi^2 + \frac{\lambda}{4} \phi^4$ (Higgs-like below). The parameter choices are fixed at $V_0,\phi_0,a,g,\lambda,=1,r,m=0.1 $, For the exponential plot the blue colour denotes negative powers of $\phi$, red colour denotes positive odd powers of $\phi$ and purple color denotes positive even powers of $\phi$}
\label{exp_evol}
\end{figure}

The evolution of the scalar field is defined by Eq. (\ref{wave2}). Using Eq. (\ref{finalexactevo}), we re-write this equation as
\begin{equation}\label{rk4waveequation}
\ddot{\phi}=\Bigg[\frac{2(2\gamma_0(t-t_0) )}{\gamma_0(t-t_0)^2-\gamma_0r^2}\Bigg]\dot{\phi}-\frac{V'(\phi)}{\Big[{\gamma_0(t-t_0)^2-\gamma_0r^2}\Big]^2}.
\end{equation}
We solve for $\phi$ numerically, taking particular values of the parameters $\gamma_0$, $r_{b}$, $r$ and choosing three different self-interaction potentials. For an exponential potential of the form $V(\phi) = a + b e^{g\phi}$, the scalar field is monotonically decreasing in nature. As shown in Fig. \ref{exp_evol}, for a logarithmic potential $V(\phi) = V_0 \log \left(1 + \frac{\phi}{\phi_0^2}\right)$, it is monotonically increasing while it is oscillatory for a Higgs-like self interaction $V(\phi) = V_0 + \frac{1}{2}m^2 \phi^2 + \frac{\lambda}{4} \phi^4$. This is a qualitatiuve nature and does not alter for different choices of parameters.

\subsection{Curvature Analysis}
The nature of singularity has been analyzed from Ricci scalar obtained from the metric function
\begin{equation}\label{exactevoconf2}
A(r,t)=\gamma_0(t-t_0)^2-\gamma_{0}r^2.
\end{equation}
It is straightforward to check that the Ricci scalar never diverges, and in fact, the Ricci scalar $R=0$. The vanishing Ricci scalar confirms that the geometry is Ricci-flat. However, the spacetime contains a nontrivial scalar field together with an effective perfect-fluid distribution and therefore, a vanishing of the Ricci scalar in this context does not imply the absence of matter. Rather, it imposes a strong algebraic restriction on the total stress-energy tensor through the trace of Einstein equations, i.e., $R = -T = - T^{\mu}_{\ \mu}$. Therefore, the condition $R=0$ immediately requires $T=0$, which indicates that the combined matter sector must behave effectively as a traceless source. For the perfect-fluid contribution,
\begin{equation}
T^{(m)}_{\mu\nu}=(\rho+p)u_{\mu}u_{\nu}-pg_{\mu\nu},
\end{equation}
the trace is $T_{m}=\rho-3p$. Similarly, for the minimally coupled scalar field,
\begin{equation}
T^{(\phi)}_{\mu\nu} = \partial_{\mu}\phi\partial_{\nu}\phi - g_{\mu\nu}
\left[\frac{1}{2}\partial_{\alpha}\phi\partial^{\alpha}\phi - V(\phi) \right],
\end{equation}
the trace becomes
\begin{equation}
T_{\phi} = -\partial_{\alpha}\phi\partial^{\alpha}\phi + 4V(\phi) = -A^{2}\dot{\phi}^{2} + 4V(\phi),
\end{equation}
since the scalar field is homogeneous. The total trace is therefore derived as
\begin{equation}
T = T_{m}+T_{\phi} = \rho - 3p - A^{2}\dot{\phi}^{2} + 4V(\phi).
\end{equation}

Thus the Ricci-flatness effectively leads to the nontrivial constraint
\begin{equation}
\rho-3p = A^{2}\dot{\phi}^{2} - 4V(\phi)
\label{traceconstraint}
\end{equation}
throughout the spacetime evolution. Eq. (\ref{traceconstraint}) shows that the fluid and scalar sectors cannot evolve independently. Instead, the geometry dynamically enforces a cancellation between their respective trace contributions. The resulting matter distribution behaves effectively as a conformal or radiation-like medium even though neither component is individually traceless.  \\

This observation is physically significant. The spacetime is conformally flat and simultaneously Ricci-flat, implying that the entire curvature structure is encoded in the traceless part of the Ricci tensor. In this sense, the geometry resembles spacetimes sourced by conformally invariant matter distributions such as pure radiation or Maxwell fields, despite the presence of a massive scalar field and an effective fluid sector. The trace constraint also provides direct information regarding the effective equation of state of the fluid component. For a massless scalar field, $V(\phi)=0$, and the scalar trace reduces to
\begin{equation}
T_{\phi}=-A^{2}\dot{\phi}^{2}<0.
\end{equation}
The condition $T=0$ then requires
\begin{equation}
\rho-3p=A^{2}\dot{\phi}^{2}>0,
\end{equation}
which implies
\begin{equation}
\rho>3p.
\end{equation}
Thus, in the absence of a self-interaction potential, the effective fluid behaves softer than a radiation fluid, in order to compensate for the negative trace contribution of the kinetic scalar field. On the other hand, in a potential-dominated regime where
\begin{equation}
V(\phi)\gg A^{2}\dot{\phi}^{2},
\end{equation}
the scalar trace becomes positive,
\begin{equation}
T_{\phi}\approx 4V(\phi),
\end{equation}
and the trace constraint yields
\begin{equation}
\rho-3p<0,
\end{equation}
or equivalently,
\begin{equation}
p>\frac{\rho}{3}.
\end{equation}
The effective fluid must therefore acquire a comparatively stiff or exotic equation of state in order to maintain the traceless nature of the total energy-momentum tensor. This behavior is qualitatively consistent with the violation of classical energy conditions observed in the fluid sector, as discussed in the next subsection.  \\

It is also worth emphasizing that the condition $R=0$ does not imply $R_{\mu\nu}=0$. The Ricci tensor remains non-vanishing and is sourced entirely by the traceless part of the total stress-energy tensor. As a consequence, the collapse dynamics is governed by a highly constrained effective matter configuration in which the scalar field and fluid components continuously redistribute their contributions so as to preserve the vanishing scalar curvature throughout the evolution.

\subsection{Energy Conditions for the Scalar Field and the Fluid}
Having solved for the conformal factor and the scalr field, it is now important to understand the behavior of the accompanying perfect fluid. For a physically viable matter distribution, it is essential that the energy-momentum tensor satisfies appropriate energy conditions \cite{Kolla2, chan2, Goswami4, pankajritu}. For example, the Null Energy Condition (NEC) requires that for any null vector $k^{\mu}$, $T_{\mu\nu} k^{\mu} k^{\nu} \geq 0$. The Weak Energy Condition (WEC) requires that for any timelike vector $w^{\alpha}$, $T_{\alpha\beta} w^{\alpha} w^{\beta} \geq 0$. This ensures that the energy density measured by any observer is non-negative. The Dominant Energy Condition (DEC) requirs that for any timelike vector $w^{\alpha}$, the vector $-T_{\alpha\beta}w^{\beta}$ must be timelike or null. Physically, this implies that the flow of energy does not exceed the speed of light. The Strong Energy Condition (SEC) requires that for any timelike unit vector $w^{\alpha}$, $2 T_{\alpha\beta} w^{\alpha} w^{\beta} + T \geq 0$, where $T$ is the trace of the energy-momentum tensor. This condition is related to the attractive nature of gravity and may be violated in the presence of sufficiently negative pressure. A detailed analysis of energy conditions for imperfect fluids can be found in literature \cite{Kolla2, pimentel}. For the present configuration, the algebraic form of these conditions is expressed as 

\begin{itemize}
\item{ (Null energy conditions; $NEC1$ and $NEC2$) 
\begin{eqnarray}\label{ec1}
&\mid \rho + p_{r} \mid- 2\, \mid q\mid \geq 0 ~,\\
&\rho - p_{r} + 2\,p_{t}+\bigtriangleup \geq 0 ~,
\end{eqnarray}
}

\item{\textit{Weak} energy condition (WEC)
\begin{equation}\label{ec2}
\rho - p_{r} +\bigtriangleup \geq 0 ~,
\end{equation}
}

\item{\textit{Dominant} energy conditions ($DEC1$ and $DEC2$)
\begin{eqnarray}\label{ec3}
&\rho - p_{r} \geq 0 ~, \\
&\rho - p_{r} -2\,p_{t} +\bigtriangleup \geq 0 ~,
\end{eqnarray}
}

\item{\textit{Strong} energy condition (SEC)
\begin{equation}\label{ec4}
2\,p_{t}+ \bigtriangleup \geq 0~,
\end{equation}
}
\end{itemize}

where $\bigtriangleup = \sqrt{(\rho + p_{r})^2 - 4\,q^2}$. To analyze the physical viability of the collapsing configuration, we examine the evolution of these conditions for a particular radial shell. The effective energy density and pressure can be obtained from the field Eqs. (\ref{fe1}) and (\ref{fe2}). Using the exact solution we find that for the effective fluid distribution
\begin{eqnarray}\nonumber
&& \rho_{m} = 12{\gamma_{0}}^2(t-t_0)^2-12r^2{\gamma_{0}}^2-12\gamma_{0}A\\&&
-\frac{1}{2}A^2\dot{\phi}^2-V(\phi),\\&&\nonumber
p_{m} = -12{\gamma_{0}}^2(t-t_0)^2+12r^2{\gamma_{0}}^2+12\gamma_{0}A\\&&
-\frac{1}{2}A^2\dot{\phi}^2+V(\phi).
\end{eqnarray}

Whether or not the energy conditions are satisfied, depends sensitively on the evolution of conformal factor and scalar field. The strong energy condition, for example, is expressed through the inequality
\begin{equation}
(\rho_m + 3p_m = 2V(\phi) -2\dot{\phi}^{2} [\gamma_0(t-t_0)^2-\gamma_{0}r^2]^2.
\end{equation}
Since $\dot{\phi}^2$ is positive definite for a canonical scalar field, the strong energy condition is only satisifed depending on the relative magnitude of $V(\phi)$, provided
\begin{equation}\label{condition}
-\frac{{V(\phi)}^{1/2}}{(\gamma_0(t-t_0)^2-\gamma_{0}r^2)} \leq \dot{\phi} \leq \frac{{V(\phi)}^{1/2}}{(\gamma_0(t-t_0)^2-\gamma_{0}r^2)}.
\end{equation}

\begin{figure}[h!]
\centering
\includegraphics[width=0.4\textwidth]{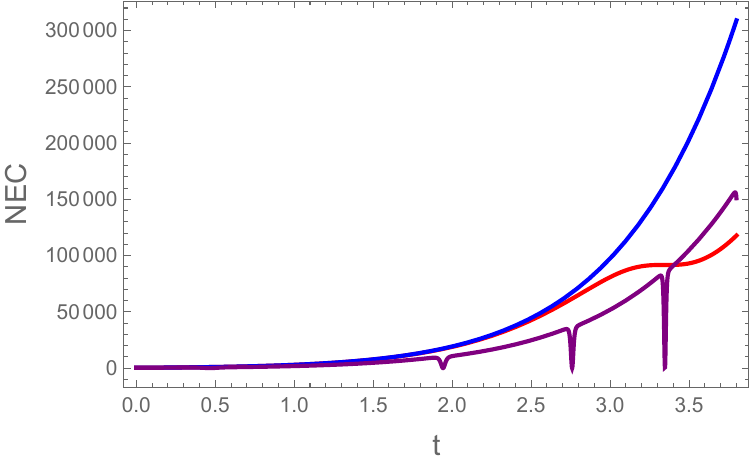}
\includegraphics[width=0.4\textwidth]{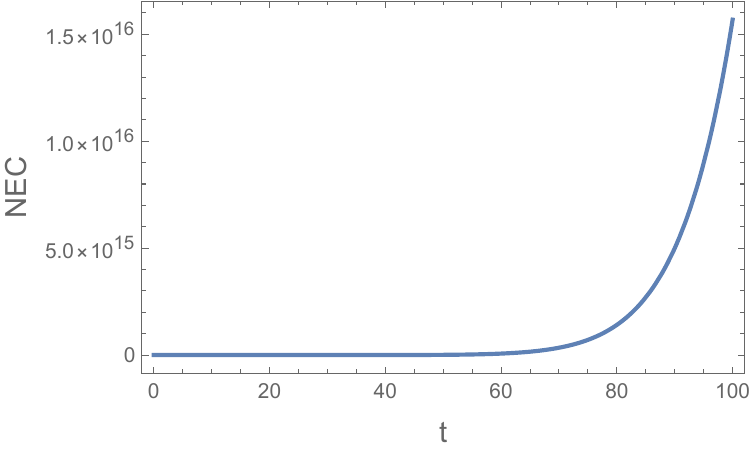}
\includegraphics[width=0.4\textwidth]{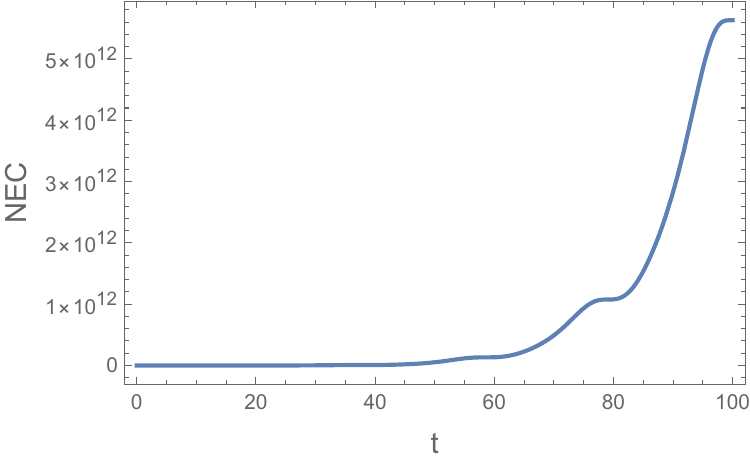}
\caption{Null Energy Condition for scalar field of three different potentials a)Exponential, b)Log, c)Higgs-field which were mentioned in Section.5, where the values for the parameters were fixed as $V_0,\phi_0,a,g,\lambda,=1,r,m=0.1 $, For the exponential plot the blue colour denotes negative powers of $\phi$, red colour denotes positive odd powers of $\phi$ and purple color denotes positive even powers of $\phi$ }
\label{nec_scalar_exp}
\end{figure}

\begin{figure}[h!]
\centering
\includegraphics[width=0.4\textwidth]{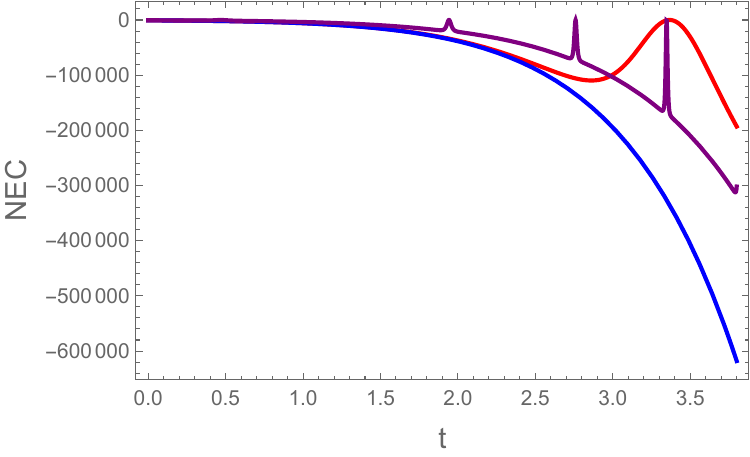}
\includegraphics[width=0.4\textwidth]{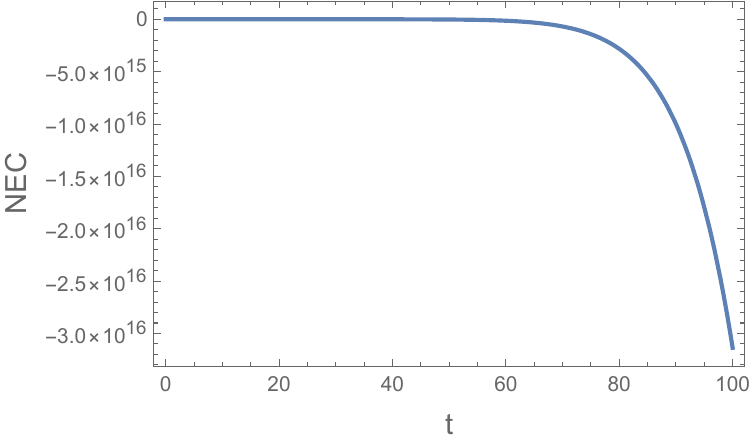}
\includegraphics[width=0.4\textwidth]{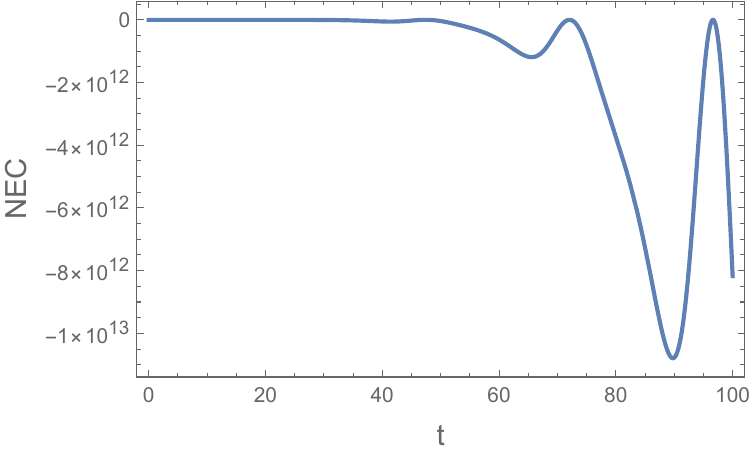}
\caption {Null Energy Condition for fluid of three different potentials a)Exponential, b)Log, c)Higgs-field which were mentioned in Section.5, where the values for the parameters were fixed as $V_0,\phi_0,a,g,\lambda,=1,r,m=0.1 $, For the exponential plot the blue colour denotes negative powers of $\phi$, red colour denotes positive odd powers of $\phi$ and purple color denotes positive even powers of $\phi$ } 
\label{nec_fluid_exp}
\end{figure}

\begin{figure}[h!]
\centering
\includegraphics[width=0.4\textwidth]{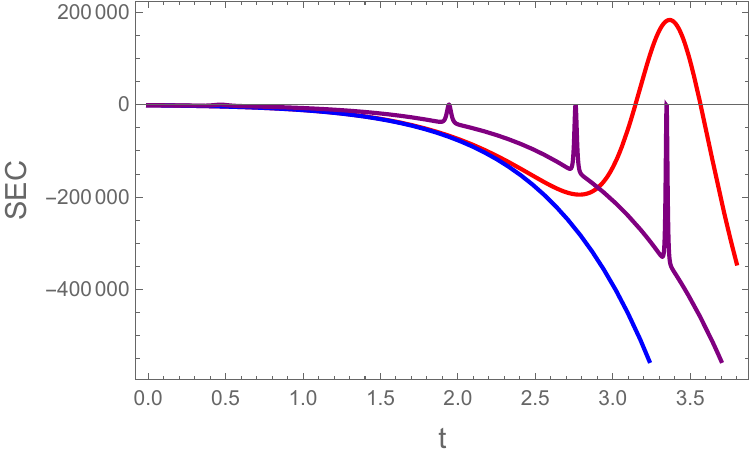}
\includegraphics[width=0.4\textwidth]{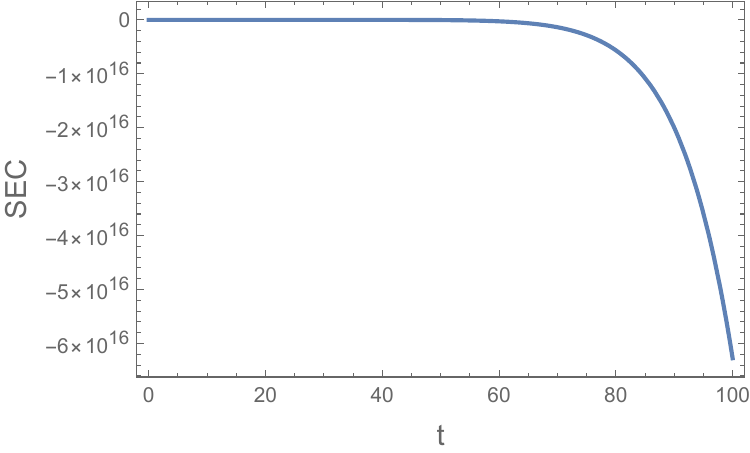}
\includegraphics[width=0.4\textwidth]{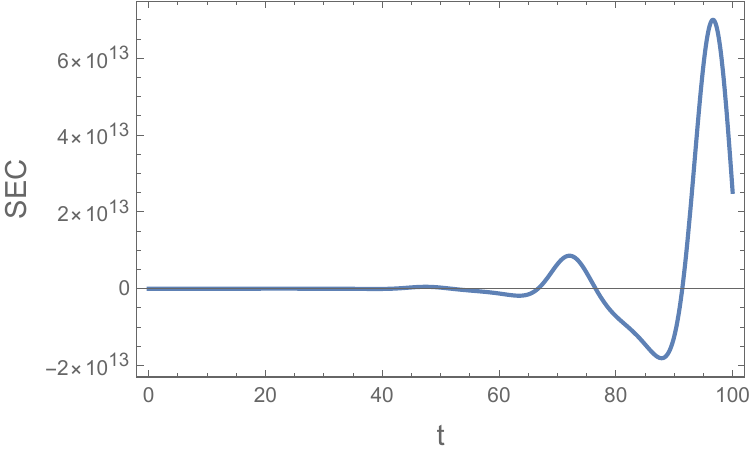}
\caption {Strong Energy Condition for fluid of three different potentials a)Exponential, b)Log, c)Higgs-field which were mentioned in Section.5, where the values for the parameters were fixed as $V_0,\phi_0,a,g,\lambda,=1,r,m=0.1 $, For the exponential plot the blue colour denotes negative powers of $\phi$, red colour denotes positive odd powers of $\phi$ and purple color denotes positive even powers of $\phi$ }
\label{sec_fluid_exp}
\end{figure}

Even when the above condition is not satisfied and we have a violation of the strong energy condition, collapse solutions can indeed exist, for suitable choices of parameters ($\gamma_{0}$ and $V_{0}$). Such scenarios can be interpreted as collapse in the presence of exotic matter, such as a clustering of dark energy \cite{tband, koyel}. Moreover, a violation of energy conditions are not entirely forbidden, for instance, corrections from quantum field theory permit temporary violations of classical energy conditions \cite{roman, ford}. Additionally, models involving negative energy densities have been explored in cosmological contexts, leading to rich dynamical behavior such as cyclic evolution and cosmological bounces \cite{nemi, noah}. Although these scenarios challenge classical thermodynamic expectations \cite{hawk}, they provide valuable insights into possible extensions of theories of gravitaty.  \\

For completeness, we analyze the energy conditions separately for the scalar field and the fluid components. NEC for the scalar field is given by $\rho_{\phi} + p_{\phi} \ge 0$. Similarly, for the fluid component, it is given by $\rho_m + p_m \ge 0$. The SEC for the fluid requires $\rho_m + 3p_m \ge 0$. We use the numerical solutions of the scalar field, for three different potentials and plot the corresponding NEC conditions for the scalar field in Fig. \ref{nec_scalar_exp}. Similarly, we plot the NEC and SEC for the fluid components in Fig. \ref{nec_fluid_exp} and Fig. \ref{sec_fluid_exp}. It is evident that the scalar field sector satisfies the NEC throughout the evolution for all the potentials considered, whereas the fluid component exhibits a clear violation of both NEC and SEC. From a physical perspective, this suggests that the collapse is not driven by a conventional perfect fluid alone, but rather by a composite system where the scalar field governs the dominant energy contribution, and the fluid behaves as an effective dark energy-like component. The violation of the strong and null energy conditions may be associated with repulsive gravitational effects, potentially delaying or altering the formation of trapped surfaces.

\section{A Self-Similar Collapse with Dissipative Flux}
In this section, we explore if a self-similar conformal factor of the form $A = A(t/r)$ can be derived from the field equations governing the gravitational collapse. The notion of self-similarity is related to the existence of Killing vectors, more precisely, conformal Killing vector fields. A vector field $X$ is said to be a Killing vector if it satisfies the Killing equation
\begin{equation}\label{killing}
L_{X}g_{ab} = g_{{ab},c}X^c + g_{cb}X^c{}_{,a} + g_{ac}X^c{}_{,b} = 0,
\end{equation}
where $L_{X}$ denotes the Lie derivative along $X$. The existence of a Killing vector is associated with a conserved quantity; for instance, a timelike Killing vector implies conservation of energy. If the Lie derivative of the metric is proportional to the metric itself, i.e., $L_{X}g_{ab}=2\Phi g_{ab}$, $X$ is referred to as a conformal Killing vector, and the spacetime admits a conformal symmetry. Such symmetries preserve angles while allowing local rescaling of distances.  $\Phi$ is a scalar function. For a constant $\Phi$, $X$ is called a homothetic Killing vector and the spacetime is said to be self-similar. A self-similar spacetime admits scale-invariant structures, where physical quantities exhibit identical behaviour under appropriate rescalings of coordinates. In spherically symmetric spacetimes, a typical homothetic vector takes the form $X^a = (t,r,0,0)$ in suitable coordinates. The existence of such a vector implies that the metric functions depend on a single dimensionless variable, usually taken as $t/r$.  \\

We investigate whether the present collapsing system admits solutions of the form $A = A(t/r)$. We find that the pressure isotropy condition as in Eq. (\ref{isotropy}), indeed allows for a self-similar solution, however, the $G_{01}$ field equation impose an additional constraint that may lead to inconsistency with just a perfect fluid. Only thorugh an inclusion of radial heat flux can modify the structure enough and accommodate a consistent self-similar solution. We write the matter content by including a radial heat flux as
\begin{equation}
T_{\mu\nu} = (\rho + p)u_\mu u_\nu - p g_{\mu\nu} + q_\mu u_\nu + q_\nu u_\mu,
\end{equation}
where $\rho$ and $p$ denote the energy density and isotropic pressure, respectively, and $q^\mu$ is the heat flux vector. In a comoving frame. The $G_{01}$ component of the field equations nnow leads to
\begin{equation}
\frac{2\dot{A}'}{A} = -8\pi q A^2.
\label{heatflux_eq}
\end{equation}
It is important to note that for $q = 0$, $\dot{A}'=0$ which does not allow one to accommodate a self-similar solution. The pressure isotropy condition retains its original form as
\begin{equation}
\frac{A''}{A} - \frac{A'}{rA} = 0.
\label{isotropy_ss}
\end{equation}

Imposing a self-similar ansatz of the form $A(r,t) = A\left(\frac{t}{r}\right)$, we solve the above equation to write
\begin{equation}
A(r,t) = D - k \frac{r^2}{t^2},
\label{ss_solution}
\end{equation}
where $D$ and $k$ are constants. For the conformally flat metric under consideration, areal radius is $Y(r,t) = \frac{r}{A(r,t)}$ and a zero proper volume singularity corresponds to a shell-focusing condition $Y(r,t) \rightarrow 0$. For a finite comoving coordinate $r$, this requires $A(r,t) \rightarrow \infty$. From Eq. (\ref{ss_solution}), we find that for finite values of $r > 0$ and $t > 0$, $A(r,t)$ does not diverge. Therefore, the condition $A \to \infty$ is never realized during the evolution, and the proper radius $Y(r,t)$ does not vanish. As a result, the spacetime does not have any shell-focusing singularity. We derive the Ricci scalar as
\begin{equation}\label{ricci1}
R=\frac{12k(-r^2+t^2)(kr^2+3 D t^2)}{t^6},
\end{equation}
which remains finite for all $t > 0$. Substituting Eq. (\ref{ss_solution}) into Eq. (\ref{heatflux_eq}), we derive the corresponding heat flux as
\begin{equation}
q(r,t) = -\frac{1}{\pi} \frac{k r}{t^3 A^3}.
\label{heatflux_solution}
\end{equation}

Similarly, using the field equations, the effective energy density and pressure can be derived as
\begin{equation}
\rho = 3\dot{A}^2 - 3A'^2 + 2AA'' + \frac{4}{r}AA' - \frac{1}{2}A^2 \dot{\phi}^2 - V(\phi),
\end{equation}
\begin{equation}
p = 2A\ddot{A} - 3\dot{A}^2 + 3A'^2 - \frac{4}{r}AA' - \frac{1}{2}A^2 \dot{\phi}^2 + V(\phi).
\end{equation}

\begin{figure}[h!]
\centering
\includegraphics[width=0.4\textwidth]{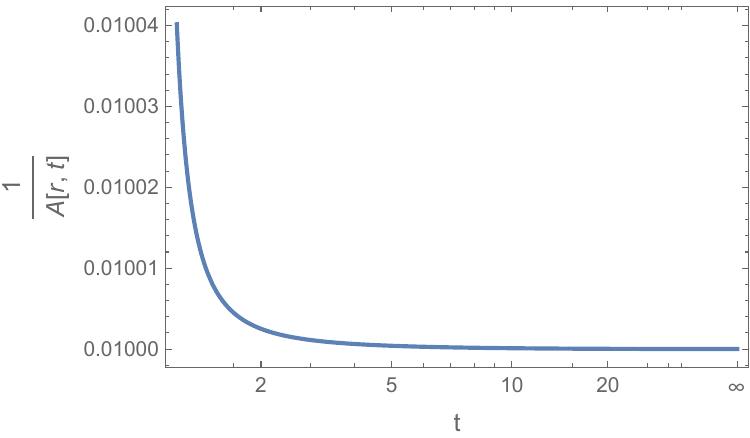}
\includegraphics[width=0.4\textwidth]{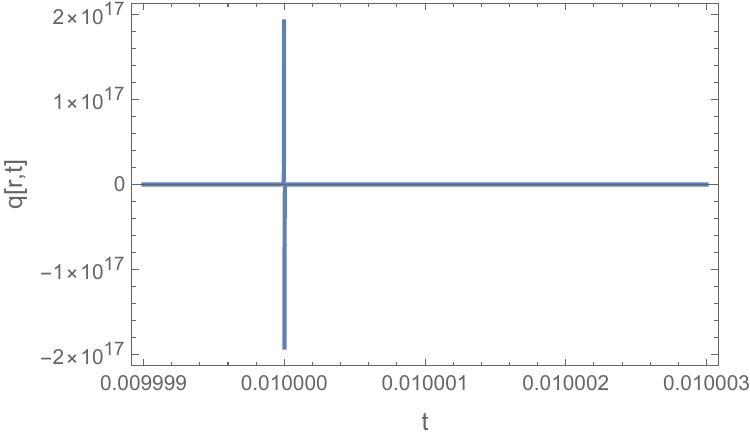}
\caption{The first image shows the Time-evolution of the scale factor $Y(r,t) \propto \frac{1}{A(r,t)}$ for specific values of $r=1$,$D=100$ and $k=0.01$; the figure shows the time evolution when $t_0 = 0$ and the second image shows the heat flux $q[r,t]$ }
\label{fig:conformal1}
\end{figure}

\begin{figure}[h!]
\centering
\includegraphics[width=0.3\textwidth]{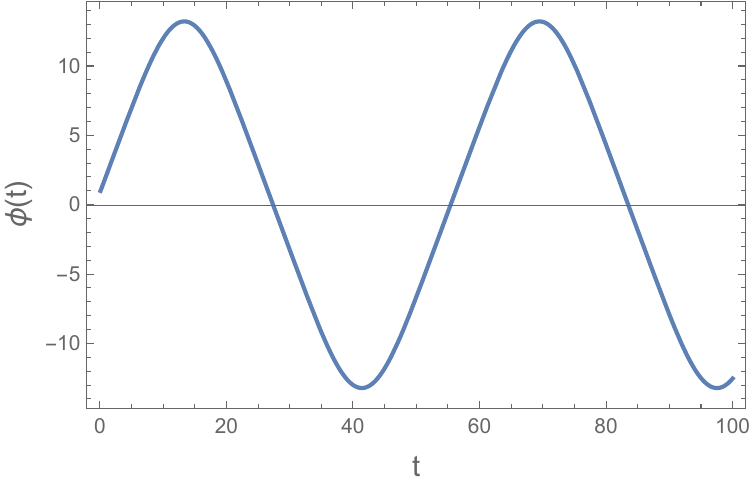}
\includegraphics[width=0.3\textwidth]{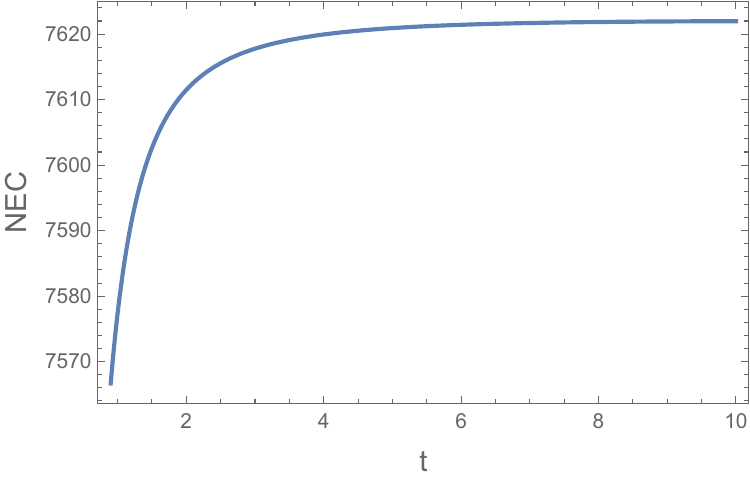}
\includegraphics[width=0.3\textwidth]{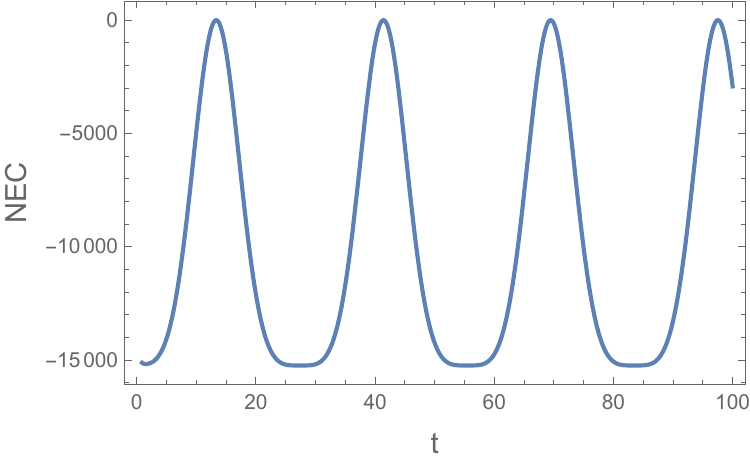}
\includegraphics[width=0.3\textwidth]{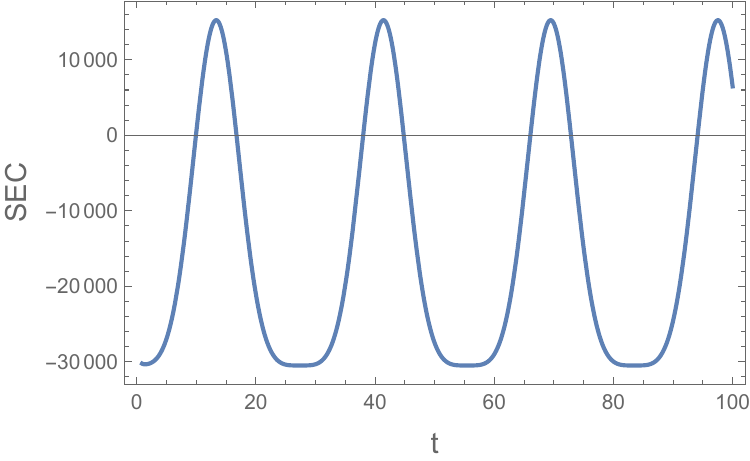}
\caption{The first plot shows the Time-evolution of the scalar field $\phi(t)$ second image shows the Null Energy Condition for scalar field third image shows the Null Energy Condition for matter fluid and fourth image shows the Strong Energy Condition for matter fluid for specific values of $r = 1$,$D = 100$ and $k = 0.01$ }
\label{fig:conformal2}
\end{figure}

Using the Klein-Gordon equation for the scalar field Eq. (\ref{wave2}), we solve for the scalar field that supports the self-similar conformal factor, following a similar strategy as in the last section. However, we choose only one example of self-interaction potential in this section, the Higgs potential. The numerical solution is shown in the top panel of Fig. \ref{fig:conformal2}. Using the solution of th scalar field, expressions of energy density, pressure and heatflux we derive and plot the Null energy condition of the scalar field (second from the top in Fig. \ref{fig:conformal2}), Null energy condition and Strong energy condition of the accompanying fluid (third and fourth from the top in Fig. \ref{fig:conformal2}). The scalar field is oscillatory in nature and this behavior is clearly manifested in the energy conditions. We find that the Null energy condition for the scalar field is always satisfied, but the fluid distribution violates both the Null and Strong energy conditions.    \\

One additional thing to consider for a self-similar case is the fact that, in the presence of a non-vanishing radial heat flux, the exterior spacetime cannot be described by a Schwarzschild solution. A consistent exterior geometry is instead given by the outgoing Vaidya spacetime \cite{vaidya}. We briefly discuss the matching of an interior conformally flat spacetime to a Vaidya exterior across a timelike hypersurface $\Sigma$ defined by $r=r_b$. The interior metric is
\begin{equation}
ds^2_- = \frac{1}{A(r,t)^2}\left(dt^2 - dr^2 - r^2 d\Omega^2\right),
\end{equation}
while the exterior Vaidya metric is
\begin{equation}
ds^2_+ = \left(1 - \frac{2M(v)}{R}\right) dv^2 + 2\, dv\, dR - R^2 d\Omega^2.
\end{equation}

The boundary $\Sigma$ is comoving in the interior and is parametrized by $r=r_b=\text{constant}$, with intrinsic coordinates $(t,\theta,\phi)$. The areal radius of the boundary is
\begin{equation}
R_\Sigma(t) = \frac{r_b}{A(r_b,t)}.
\label{Rboundary}
\end{equation}

The induced metric on $\Sigma$ from the interior is
\begin{equation}
ds^2_{\Sigma^-} = \frac{1}{A(r_b,t)^2} dt^2 - R_\Sigma^2(t) d\Omega^2.
\end{equation}

From the exterior, restricting to $\Sigma$ and writing $R=R_\Sigma(t)$ and $v=v(t)$, we obtain
\begin{equation}
ds^2_{\Sigma^+} =
\left[\left(1 - \frac{2M(v)}{R_\Sigma}\right)\dot v^2 + 2 \dot v \dot R_\Sigma \right] dt^2
- R_\Sigma^2 d\Omega^2.
\end{equation}

Matching the angular components immediately yields Eq. (\ref{Rboundary}). Matching the temporal components gives
\begin{equation}
\frac{1}{A^2} = \left(1 - \frac{2M}{R_\Sigma}\right)\dot v^2 + 2 \dot v \dot R_\Sigma.
\label{ffmatch}
\end{equation}

Using $\dot R_\Sigma = -\frac{r_b \dot A}{A^2}$ and Eq. (\ref{ffmatch}) one can determine $\dot v$ implicitly. The extrinsic curvature components can be derived by using the unit normal to $\Sigma$. The outward unit normal to $\Sigma$ is $n_\mu^- = \left(0, \frac{1}{A}, 0, 0\right)$, which satisfies $n^\mu n_\mu = -1$. The extrinsic curvature is defined as
\begin{equation}
K_{ab}^- = - n_{\mu;\nu} e^\mu_a e^\nu_b.
\end{equation}

For the angular component, we find that
\begin{equation}
K_{\theta\theta}^- = - n_r \Gamma^r_{\theta\theta}.
\end{equation}

For the interior metric,
\begin{equation}
\Gamma^r_{\theta\theta} = -r A^{-2} + r^2 \frac{A'}{A^3},
\end{equation}
which gives
\begin{equation}
K_{\theta\theta}^- = \frac{r}{A^3} - \frac{r^2 A'}{A^4}.
\end{equation}

Evaluated at $\Sigma$,
\begin{equation}
K_{\theta\theta}^- \big|_\Sigma = \frac{r_b}{A^3} - \frac{r_b^2 A'}{A^4}.
\label{Kminus_final}
\end{equation}

For the exterior, the boundary is described by $R=R_\Sigma(t)$ and $v=v(t)$. The normal to $\Sigma$ is
\begin{equation}
n_\mu^+ = (-\dot R_\Sigma, \dot v, 0, 0),
\end{equation}
up to a normalization. The angular component of the extrinsic curvature is
\begin{equation}
K_{\theta\theta}^+ = - R_\Sigma \left( \dot R_\Sigma + \left(1 - \frac{2M}{R_\Sigma}\right)\dot v \right).
\label{Kplus_final}
\end{equation}

The Darmois-Israel condition requires $K_{\theta\theta}^- = K_{\theta\theta}^+$. Using Eqs.~(\ref{Kminus_final}) and (\ref{Kplus_final}) along with $R_\Sigma = r_b/A$, and substituting $\dot R_\Sigma$, the above reduces  $p|_\Sigma = q|_\Sigma$, which is the standard condition ensuring continuity of momentum flux across the boundary. The Misner-Sharp mass is defined as
\begin{equation}
m(t,r) = \frac{R}{2}\left(1 + g^{\mu\nu}\partial_\mu R \partial_\nu R \right),
\end{equation}
with $R=r/A$. This yields
\begin{equation}
m(t,r) = \frac{r}{2A} \left[\frac{r^2}{A^2}(\dot{A}^2 - A'^2) + \frac{2r A'}{A}\right].
\end{equation}

The matching condition requires $M(v) = m(t,r_b)$, which decreases with time due to the presence of heat flux. The relation between $v$ and $t$ follows from Eq.~(\ref{ffmatch}). Substituting $R_\Sigma = r_b/A$ and $\dot R_\Sigma$, we obtain
\begin{equation}
\frac{1}{A^2} = \left(1 - \frac{2M}{R_\Sigma}\right)\dot v^2 - 2 \dot v \frac{r_b \dot A}{A^2}.
\end{equation}

This is a quadratic equation in $\dot v$, yielding
\begin{equation}
\dot v = \frac{\frac{r_b \dot A}{A^2} \pm \sqrt{\left(\frac{r_b \dot A}{A^2}\right)^2 + \left(1 - \frac{2M}{R_\Sigma}\right)\frac{1}{A^2}}}
{\left(1 - \frac{2M}{R_\Sigma}\right)}.
\end{equation}

Thus, $v(t)$ is obtained by integration, and the exterior mass can be expressed parametrically. Overall, this matching demonstrates that the interior self-similar solution represents a radiating configuration consistently embedded in a Vaidya exterior. The mass function decreases due to outward energy flux, and the absence of a static exterior reflects the dissipative nature of the collapse.

\section{Conclusion}

In this work, we have investigated the gravitational collapse of a massive scalar field in a conformally flat, spherically symmetric spacetime, considering both non-dissipative and dissipative matter configurations. The assumption of conformal flatness, implemented through the vanishing of the Weyl tensor, imposes strong geometric restrictions on the spacetime while still allowing for non-trivial collapse dynamics. This framework enables the construction of exact analytical solutions and provides a useful setting for examining the interplay between scalar field dynamics, symmetry properties, and dissipation during collapse. \\

For the non-dissipative configuration, the pressure isotropy condition leads to a separable conformal factor and consequently to an exact collapsing solution matched smoothly to a Schwarzschild exterior. The resulting evolution describes a continuously contracting configuration whose proper radius decreases asymptotically without reaching a shell-focusing singularity within finite proper time. The collapse therefore remains eternally ongoing from the perspective of the comoving observer. This behavior indicates that the conformal structure of the spacetime can significantly influence the late-time evolution of collapse and delay the onset of singular behavior. \\

We further explored the existence of self-similar solutions associated with homothetic symmetry. Although the isotropy condition formally admits self-similar conformal factors, the complete Einstein field equations impose an additional dynamical restriction that prevents their realization in the presence of a perfect fluid alone. This reveals a non-trivial incompatibility between self-similarity, isotropic matter distributions, and homogeneous scalar field evolution within the present framework. However, the introduction of dissipative effects through a radial heat flux modifies the mixed field equation and restores consistency of the self-similar ansatz. The resulting configurations possess non-vanishing outward energy transport and must therefore be matched to a radiating Vaidya exterior. The associated Misner--Sharp mass decreases monotonically during the evolution, reflecting the dissipative nature of the collapse. Importantly, even in this dissipative and self-similar regime, the conformal factor remains finite throughout the evolution and the proper radius never vanishes at finite time, indicating the absence of shell-focusing collapse within the domain of the solution. \\

The analysis of the energy conditions provides additional insight into the physical nature of the matter content. While the scalar field sector remains well behaved and satisfies the null energy condition for the potentials considered, the effective fluid component exhibits violations of the null and strong energy conditions during certain stages of evolution. Such violations may be interpreted as effective exotic behavior arising from the redistribution of energy between the scalar and fluid sectors, generating negative-pressure contributions reminiscent of dark-energy-like matter. Although classical energy conditions are violated in these regimes, similar features commonly arise in scalar-field cosmology, semiclassical gravity, and models involving dissipative or quantum corrections. The violation of these conditions may also influence the development of trapped surfaces and modify the causal structure of the collapsing spacetime. \\

Overall, the present analysis demonstrates that conformal symmetry, scalar field dynamics, and dissipative transport can collectively produce qualitatively different collapse outcomes compared to conventional perfect-fluid collapse scenarios. In particular, dissipative effects appear to play a crucial role in enabling self-similar evolution while simultaneously altering the approach toward singular configurations. The solutions obtained here therefore provide a useful analytical framework for exploring non-singular or asymptotically collapsing scalar-field systems in general relativity. \\

There remain several important directions for future investigation. It would be worthwhile to analyze curvature invariants and trapped surface formation in greater detail in order to better characterize the global structure and singularity properties of the spacetime. Extensions involving spatially inhomogeneous scalar fields, anisotropic pressures, more general self-interaction potentials, or modified theories of gravity may further enrich the collapse dynamics. Stability analyses of the present solutions under perturbations could also provide important insight into their physical viability and astrophysical relevance.

\end{document}